\documentclass[10pt,journal]{IEEEtranTCOM}

\usepackage[english]{babel}
\usepackage[usenames]{color}
\usepackage[cp1250]{inputenc}
\usepackage{amsfonts}
\usepackage{amsthm}
\usepackage{graphicx}
\usepackage{epsfig}
\usepackage{mathrsfs}
\usepackage{amsmath}
\usepackage{algorithm}
\usepackage{algorithmic}
\usepackage{hyperref}

\theoremstyle{plain}

\begin{document}
\newcommand{\bea}{\begin{eqnarray}}
\newcommand{\eea}{\end{eqnarray}}
\newcommand{\be}{\begin{equation}}
\newcommand{\ee}{\end{equation}}
\newcommand{\beas}{\begin{eqnarray*}}
\newcommand{\eeas}{\end{eqnarray*}}
\newcommand{\bs}{\backslash}
\newcommand{\bc}{\begin{center}}
\newcommand{\ec}{\end{center}}
\def\SC {\mathscr{C}}

\title{Modelling bid-ask spread conditional distributions\\using hierarchical correlation reconstruction}
\author{\IEEEauthorblockN{Jaros{\l}aw Duda$^1$} \IEEEauthorblockN{Robert Syrek$^2$} \IEEEauthorblockN{Henryk Gurgul$^3$}\\
\IEEEauthorblockA{$^1$ Institute of Computer Science, Jagiellonian University, Krak\'ow, Poland. Email: \emph{jaroslaw.duda@uj.edu.pl}}\\
\IEEEauthorblockA{$^2$ Institute of Economics, Finance and Management, Jagiellonian University, Krak\'ow, Poland}\\
\IEEEauthorblockA{$^3$  Department of Applications of Mathematics in Economics, Faculty of Management, AGH University
of Science and Technology, ul. Gramatyka 10, 30-067 Krak\'ow, Poland}\\
}
\maketitle

\begin{abstract}
While we would like to predict exact values, available incomplete information is rarely sufficient - usually allowing only to predict conditional probability distributions. This article discusses hierarchical correlation reconstruction (HCR) methodology for such prediction on example of usually unavailable bid-ask spreads, predicted from more accessible data like closing price, volume, high/low price, returns. In HCR methodology we first normalize marginal distributions to nearly uniform like in copula theory. Then we model (joint) densities as linear combinations of orthonormal polynomials, getting its decomposition into (mixed) moments. Then here we model each moment (separately) of predicted variable as a linear combination of mixed moments of known variables using least squares linear regression - getting accurate description with interpretable coefficients describing linear relations between moments. Combining such predicted moments we get predicted density as a polynomial, for which we can e.g. calculate expected value, but also variance to evaluate uncertainty of such prediction, or we can use the entire distribution e.g. for more accurate further calculations or generating random values. There were performed 10-fold cross-validation log-likelihood tests for 22 DAX companies, leading to very accurate predictions, especially when using individual models for each company as there were found large differences between their behaviors. Additional advantage of the discussed methodology is being computationally inexpensive, finding and evaluation a model with hundreds of parameters and thousands of data points takes a second on a laptop.
\end{abstract}
\section{Introduction}

\begin{figure}[t!]
    \centering
        \includegraphics{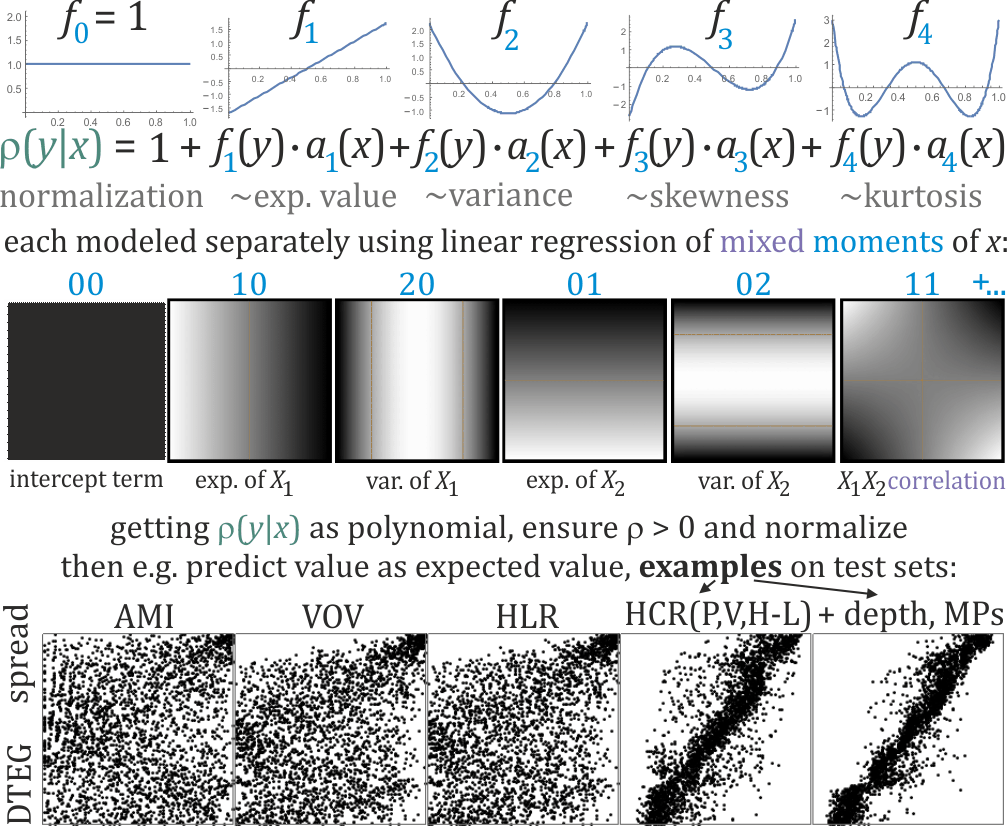}
        \caption{General concept, some firsts functions of the used 1 and 2 dimensional basis of orthornormal polynomials $(f_{j_1 j_2}(x)=f_{j_1}(x_1)f_{j_2}(x_2))$, and application example. For simplicity we assume working on variables normalized to nearly uniform marginal densities on $[0,1]$ as e.g. in copula theory. We would like to model distortion from this uniform distribution for predicted variable $Y$ based on the context $X$: as a linear combination e.g. of orthornormal polynomials here, for which coefficients have similar interpretation as moments/cumulants: $a_1$ shifts right/left like expected value, $a_2$ increases/decreases probability of extreme values as variance etc. Each such coefficient is separately modelled using analogous coefficients of $X$ variable: $\tilde{\rho}(y|x)= \sum_j f_j(y) \sum_k \beta_{jk} f_k(x)$, e.g. using least-squares linear regression here. Such predicted density as polynomial sometimes gets below zero, hence there is used e.g. $\rho=\max(\tilde{\rho},0.03)/N$ instead, with $N$ normalization factor to integrate to 1. Example of application of such predicted density is just taking its expected value, getting a conservative prediction of value (avoiding extremes), also with estimated uncertainty if additionally calculating variance of the predicted density.}
       \label{intr}
\end{figure}

\begin{figure}[t!]
    \centering
        \includegraphics{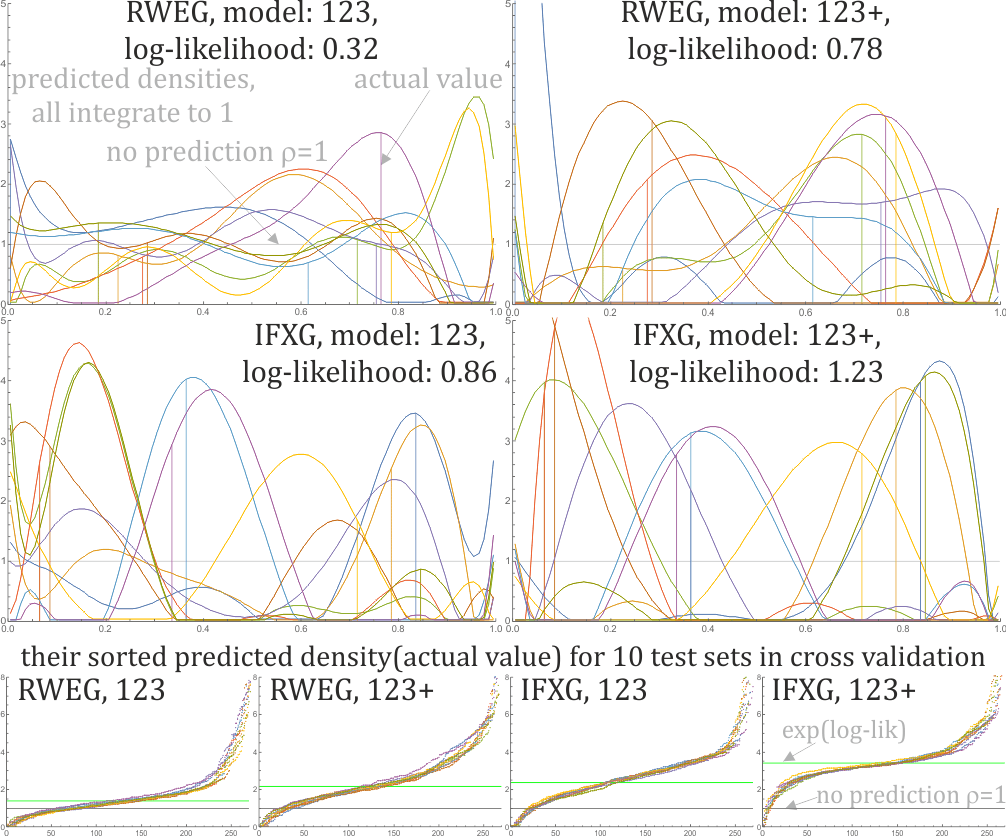}
        \caption{Top: examples of predicted conditional densities - predicted $\tilde{\rho}(y|x^i)= \sum_j f_j(y) \sum_k \beta_{jk} f_k(x^i)$ polynomial for $i$-th datapoint undergoes $\rho=\max(\tilde{\rho},0.03)/N$ to remove negative densities, and normalization to integrate to $1=\int_0^1 \rho(y|x)dy$. Each diagram contains 10 example predictions, vertical lines show the actual values $(y^i, \rho(y^i|x^i))$: the higher the better prediction, without prediction all would have height 1. There were chosen companies having best/worst prediction. The best ones predict mainly narrow unimodal distributions in line with the actual values, weaker ones can rather only predict wide often multimodal distributions. We can see rapid growths at the ends - they are likely artifacts of using polynomials, their additional removal might improve prediction. Bottom: their sorted predicted densities in the actual values $\{\rho(y^i|x^i)\}_i$, with marked gray $\rho=1$ line of using no prediction and green exp(log-likelihood) line corresponding to average improvement over no prediction. The points are of different colors denoting one of 10 rounds of 10-fold cross-validation. }
       \label{exdens}
\end{figure}

\begin{figure}[t!]
    \centering
        \includegraphics{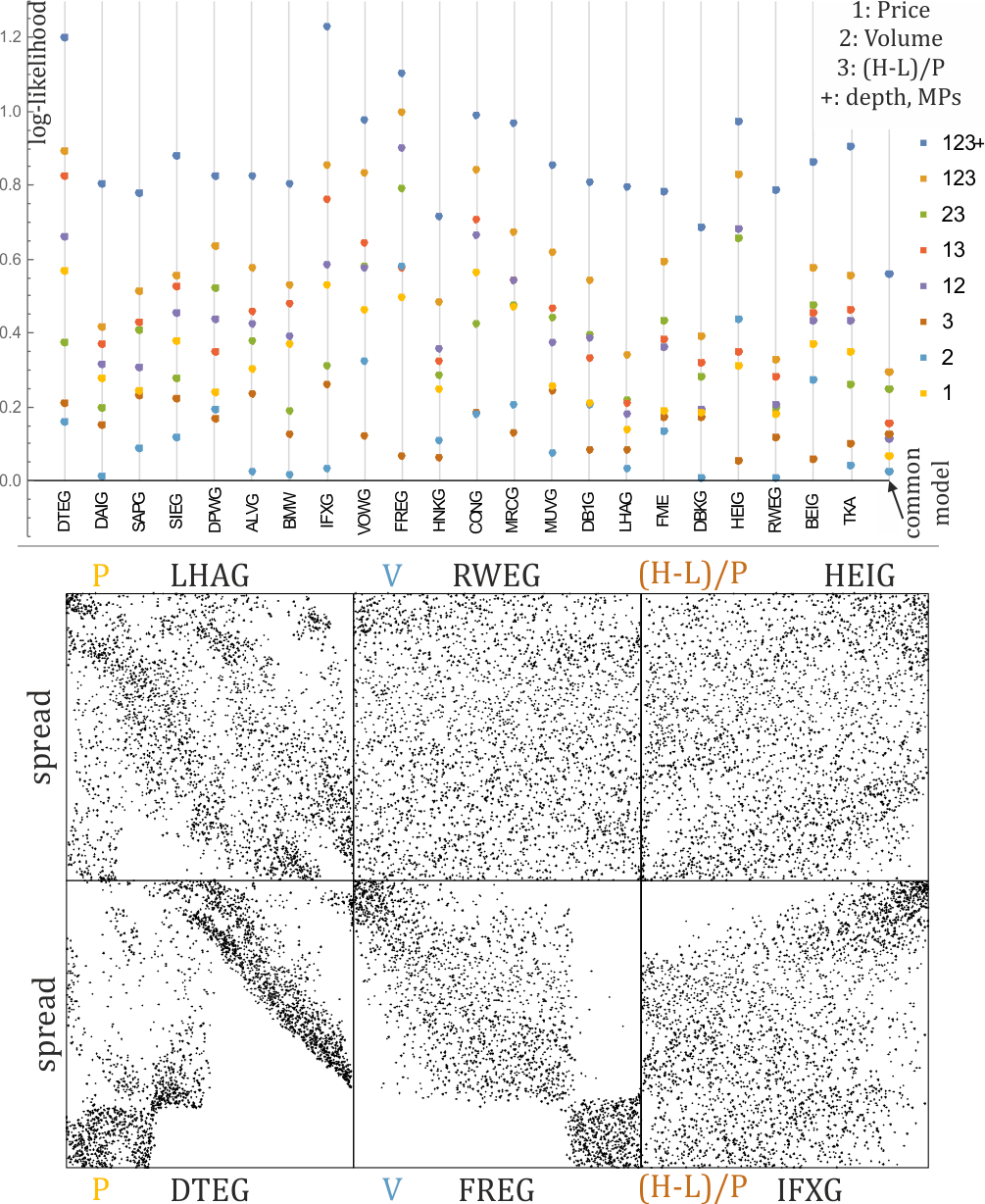}
        \caption{Top: Log-likelihoods from 10-fold cross-validation for individual models for companies using various types of information, e.g. '123' denotes using all basic 1,2,3 variables, where '1' denotes closing price ($P$), '2' volume ($V$), and '3' difference between high and low price normalized by dividing by closing price: $(H-L)/P$. The '+' denotes using also 3 additional variables: depth, midpoint changes intraday and midpoint volatility.
        The last column presents averaged evaluation for using common model for all data.
        Bottom: examples of pairwise dependencies in dataset for the 3 variables (columns) for the least and the most dependent companies for a given variable (heights of corresponding dots). For example volume ('2') does not help for RWEG (nearly uncorrelated - blue dot is near zero), but is useful for FREG. We can see that there are large differences between companies, hence we will mostly focus on building individual models for each company.
        }
       \label{lls}
\end{figure}

While it is more convenient to work on exact values, real life predictions usually have some uncertainty, controlling of which could allow e.g. to distinguish nearly certain predictions from the practically worthless ones. Generally, wanting to predict $Y$ variable from $X=(X_1,\ldots, X_d)$ variables, if there is no a strict relation, they often come from some complicated joint probability distribution - knowing $X=x$, we can only predict $\textrm{Pr}(Y|X=x)$ conditional probability distribution. This article discusses such prediction of conditional probability distributions on example of bid-ask spreads, which is often publicly unavailable, from a few variables which are available. There is used hierarchical correlation reconstruction (HCR)~\cite{HCR} methodology as briefly presented in Fig. \ref{intr}: first normalize all variables to nearly uniform distribution like in copula theory~\cite{copula}, then model densities as polynomials using basis of orthonormal polynomials - for which coefficients are analogous to (mixed) moments. Then predict such coefficients of density of $Y$ from coefficients of $x$, for example using least-squares linear regression here, separately for each modelled moment of $Y$. The evaluation is performed using 10-fold cross-validation for log-likelihood of normalized variables: average natural logarithm of predicted conditional density in the actual value, what can be interpreted as estimated minus conditional entropy $E_{XY}\left(\ln(\rho(Y|X))\right)=-H(Y|X)$. Figure \ref{exdens} contains examples of such predicted densities, Figure \ref{lls} contains main evaluation of the used variables and models.

Having predicted conditional probability distributions, a basic application can be just taking expected values - getting conservative predictions of values, e.g. avoiding predicting extremes, as presented in Figure \ref{comp}. We can additionally calculate variance of such predicted density to estimate uncertainty of value predicted as expected value. We can also handle more sophisticated situations like bimodal distribution with two (or more) separate maxima: when pointing expected value might not be a good choice (can have much lower density), a better prediction might be e.g. one of the maxima, or maybe both: providing prediction as alternative of two (or more) possibilities. Finally, we can also work on complete predicted densities, e.g. to generate its random values for Monte-Carlo methods, or processing such entire probability distributions through some further functions for their more accurate modelling, as e.g. generally $E(f(X))\neq f(E(X))$ for nonlinear functions: expected value of function is not equal function of expected value.\\

To model such conditional distributions we will use HCR methodology, which combines advantages of classical statistics and machine learning. While the former allows for well controlled and interpretable but relatively small (rough) models/descriptions, machine learning allows for very accurate descriptions using huge models, but usually lacks uniqueness of solution, control and interpretability of coefficients, and often is computationally costly. HCR allows to inexpensively work on huge models obtained from (unique) least-squares optimization, using well interpretable coefficients: as mixed moments of variables, starting e.g. with moments of single variables and correlations of coefficients.

More specifically, HCR conveniently starts with normalization of all marginal distributions to nearly uniform distributions like in copula theory - so they can be interpreted as quantiles, compactifying tails problematic for linear regression. Now we can model distortion from uniform distribution on this $[0,1]^d$ hypercube with a linear combination, e.g. of orthonormal polynomials, for which coefficients can be interpreted analogously to (mixed) moments. E.g. for 3 variables, '000' coefficient is always 1 - corresponding to normalization, '100' is analogous to expected value of 1st variable, '020' to variance of 2nd variable, '011' to correlation coefficient between 2nd and 3rd variable, '202' is large if with large variance of 1st variable there comes large variance of 3rd variable (like heteroskedasticity in ARCH model), and so on also for higher moments and dependencies between 3 or more variables, getting hierarchical decomposition of statistical dependencies (joint distribution) into mixed moments.

While we could directly extract and exploit $(X,Y)$ joint distribution with HCR, experimental tests have shown that alternative approach from \cite{cred} gives slightly better evaluation by extracting additional dependencies, hence we will focus only on it: use separate bases of (mixed) moments for $Y, X$ and predict each considered coefficient for $Y$ with least squares linear regression using coefficients of $X$. While \cite{cred} has used only moments of separate variables for $X$, here we expand this methodology by using also their mixed moments - starting with '11' correlation-like coefficient.

Its advantage over modelling joint distribution is being able to notice and exploit e.g. that difference of two variables has some useful relation with the predicted variable. Looking e.g. at RWEG in Fig. \ref{lls}, '2'-nd variable is practically noise, its (blue) dot is nearly zero. However, the difference between '13' and '123' model (red and orange dot) is much larger: relations with other variables allowed to extract more information from this noise.\\

In the discussed example we would like to predict conditional probability distributions for (nearly inaccessible) bid-ask spreads (relative quoted) from more available information. The basic considered '123' model began with 5 classically used variables: closing price ($P$), high and low value ($H,L$), volume ($V$) and log-return ($R$). Surprisingly, it has turned out that using $R$ and $L$ alone did not help improving evaluation (log-likelihood in 10-fold cross-validation), hence finally there were used 3 variables $P,V, (H-L)/P$. The second considered '123+' model complements this information with other relatively available variables, searching through which has lead to final use of 3 additional variables: (market) depth, midpoint changes intraday, and midpoint volatility.

The choice of basis of moments is a difficult question: too small leads to underfitting by not being able to express dependencies of data behavior, too large leads to overfitting by representing features of training set which do not generalize to test set. For '123' and '123+' models there was chosen a compromise to optimize for all companies: for '123' predicting 8 first moments of $Y$ using 53 mixed moments of 3 variables of $X$, hence using $8\cdot 53=424$ coefficient models. For '123+' predicting 6 moments from 205 mixed moments of 6 variables of $X$, getting  $6\cdot 205=1230$ coefficient models.

The used data is for 22 DAX companies for which large enough dataset was available, arbitrarily chosen as containing at least 2000 daily datapoints. As there were observed large difference between models for different companies, corresponding to different behaviors of traders of its stock, there were mainly considered individual models for each company. There was also performed hierarchical search for combinations of companies for which using common model leads to the smallest loss of evaluation, however, such loss often turns out significant.

\section{Dataset and basic concepts}
This section discusses dataset and reminds standard concepts, to be used for building the used methodology in the next Section.
\subsection{Dataset and variables}
There was used daily data for DAX companies from 1999-2013 period (source in Acknowledgment), selected as having available at least 2000 datapoints: Deutsche Telekom (DTEG), Daimler (DAIG), SAP (SAPG), Siemens (SIEG), Deutsche Post (DPWG), Allianz (ALVG), Bay Motoren (BMW), Infineon (IFXG), Volkswagen (VOWG), Fresenius (FREG), Henkel (HNKG), Continental (CONG), Merck (MRCG), Muench. Rueckvers (MUVG), Deutsche Boerse (DB1G), Lufthansa (LHAG), Fresen Med Care (FME), Deutsche Bank (DBKG), Heidelbergcement (HEIG), RWE (RWEG), Beiersdorf (BEIG), Theyssenkrupp (TKA).

The basic set of variables is $P$ - closing price, $V$ - volume, $R$ - return, $H, L$ - high/low price. However, it has turned out that trying to exploit dependence on $R$ and $L$ alone did improve evaluation, hence finally the basic considered model: '123' uses only $P$ as '1'-st variable, $V$ as '2'nd-variable and normalized $(H-L)/P$ as '3'-rd variable.

There were also performed trials to improve the prediction by using information from some additional relatively available variables - 3 were found helpful in predictions: (market) depth, midpoint changes intraday and midpoint volatility.
\subsection{Bid-ask spread and some its standard predictors}
Bid-ask spread is the difference between the lowest asking price ($ask$, offered by a seller) and the highest bid price ($bid$, offered by a buyer).
While this value is important because it is a main measure of market quality~\cite{liq1,liq2}, this information is usually publicly unavailable. Therefore, there is an interest in being able to predict this value based on other, more accessible data.

More specifically, we work on relative quoted spread, which is normalized by dividing by midpoint $(ask+bid)/2$:

\be S=\frac{ask-bid}{(ask+bid)/2} \ee

Simple examples of its predictor based on the 5 basic variables are $AMI$~\cite{AMI,AMI1}, 
$HLR$~\cite{HLR}:
\be AMI = \ln\left(1+\frac{|R|}{P\cdot V}   \right) \ee
\be HLR = 2\frac{H-L}{H+L} \ee
They are intended for a simpler task than discussed: to predict values, while here we want to predict entire conditional probability distributions. We can reduce predicted probability distributions into predicted values e.g. as expected value, median, or positions of maxima (especially for multimodal distributions). Figure \ref{comp} presents comparisons using such predictions reduced with expected value.

However, in practice such prediction is often further processed through some functions, generally $E(f(X))\neq f(E(X))$ for nonlinear, hence it is more accurate to process the probability distribution (e.g. on a lattice) through the functions before e.g. taking expected value.
\subsection{Normalization to nearly uniform marginal distributions}
Like in copula theory, in HCR methodology it is convenient to initially normalize all variables to nearly uniform marginal distributions in  $[0,1]$, hence we further only work on such normalized variables, what beside usually better prediction, also allows for better presentation of evaluation: e.g. density without prediction is 1, log-likelihood is 0.

This standard normalization requires estimations of cumulative distribution function (CDF), individually for each variable, and applying this CDF function to the original values. Finally, having a prediction we can go back to the original variable using CDF$^{-1}$, for example as in the original \cite{cred} article, however, for simplicity we omit this step here - working only on normalized variables. Also $AMI, HLR$ predictions underwent such normalization for the purpose of Fig. \ref{comp} visual performance comparison - making that an ideal predictor would give diagonal.

There was used empirical distribution function (EDF) for such normalization here: for each variable its $n$ observed values are sorted, then $i$-th value in such order obtains $(i-0.5)/n$ normalized value. Hence values become their estimated quantiles this way, difference of two normalized values describes percentage of population between these two values.

Having predicted density for normalized variable, we can transform it to the original variable e.g. by discretizing this density to probability distribution on $\{(i-0.5)/n\}_{i=1..n}$ lattice, and assigning probability of its $i$-th position to $i$-th ordered original value. For simplicity it is omitted in this article.

\subsection{Evaluation: log-likelihood with 10-fold cross-validation}
The most standard evaluation of probability distributions is log-likelihood like in ML estimation: average (natural) logarithm of (predicted) density in the actually observed value. Hence we will use this evaluation here.

Working on variables normalized to $\rho\approx 1$ marginal distributions, without prediction they would have practically zero log-likelihood. It allows to imagine gains from predictions as averaged improvement over this  $\rho\approx 1$, as in Fig. \ref{exdens}. For example the best observed log-likelihood $\approx 1.2$ corresponds to $\approx\exp(1.2)\approx 3.3$ times better density than without prediction, the same as if we could squeeze $[0,1]$ range 3.3 times to a $0.3$ wide range. Sorting predicted densities in the actually observed values, we can get additional information about distribution of prediction, as presented in this Figure.

We predict here conditional density - denoted as $\rho(Y=y|X=x)$ for density of $Y$ predicted based on known value of $X$. Hence the used evaluation can be seen as estimation of $E_{XY}(\ln(\rho(Y|X))$, which is minus conditional entropy $-H(Y|X)$. While being unknown, random variables have some concrete value of conditional entropy - we can hopefully try to approach it with better and better models.

We are focusing here on large models using hundreds of coefficients, hence we need to be careful not to overfit: represent only behavior which indeed generalizes, is not just a statistical artifact of the training set. Machine learning also builds large models, usually evaluating using cross-validation: randomly split dataset into training and test set, training set is used to build the model, then test (or validation) set is used to evaluate the built model.

However, such evaluation depends on the random splitting into training and test set. There is used standard 10-fold cross validation to weaken this random effect: dataset is randomly split into 10 nearly equal size subsets, evaluation is average from 10 cross validations: using successive subsets as the test set and the remaining as the training set. However, there is still observed scale $\approx 0.01$ randomness of such evaluation, hence only two digits after coma are being presented.

\section{Used HCR-based methodology}
This section discusses the used methodology, being expansion of the one used in \cite{cred}. We decompose $X$ and $Y$ variables into mixed moments and model separately each moments of $Y$ using least-squares linear regression of moments of $X$, then combine them into predicted conditional probability distribution of $Y$.
\subsection{Decomposing joint distribution into mixed moments}
After normalization of marginal distributions of all variables to nearly uniform on $[0,1]$, for $d$ variables their joint distribution on $[0,1]^d$ would be also nearly uniform if they were statistically independent.

Distortion from uniform joint distribution corresponds to statistical dependencies between these variables - we would like to model and exploit it. In HCR we model it as just a linear combination using an orthornormal basis e.g. of polynomials, which gives the coefficients similar interpretation as moments and mixed moments: dependencies between moments for multiple variables.

The first orthonormal polynomials (rescaled Legendre) for $[0,1]$ are $f_0=1$ and $f_1,f_2,f_3,f_4$ correspondingly (plotted in Fig. \ref{intr}):
$$\sqrt{3}(2x-1), \sqrt{5}(6x^2-6x+1), \sqrt{7}(20x^3-30x^2+12x-1),$$
$$3(70x^4-140x^3+90x^2-20x+1) $$
We could alternatively use e.g. $1$, $\sqrt{2}\sin(2\pi xk)$, $\sqrt{2}\cos(2\pi xk)$ for $k\geq 1$ orthonormal basis, however, experimentally it usually leads to inferior evaluation.

Decomposing density $\rho(x)=\sum_j a_j f_j(x)$, we need $a_0=1$ normalization to integrate to 1. Due to orthogonality, $\int_0^1 f_j(x)dx=0$ for $j>0$, hence the following coefficients do not affect normalization. As we can see in their plots in Fig. \ref{intr}, positive $a_1$ shifts density toward right - acting analogously as expected value. Positive $a_2$ increases probability of extreme values at cost of central values - analogously as variance. Skewness-like higher order asymmetry is brought by $a_3$ and so on - we can intuitively interpret these coefficients as moments (cumulants). This is only an approximation, but useful for interpretation of discussed models.

In multiple dimensions we can use product basis:
\be f_{j}(x)=f_{j_1}(x_1)\cdot \ldots f_{j_d}(x_d)\qquad \textrm{for }j=(j_1,\ldots,j_d)\ee
leading to model of joint distribution:
\be \rho(x) = \sum_{j\in B} f_j(x) =\sum_{j\in B} a_j f_{j_1}(x_1)\cdot \ldots\cdot f_{j_d}(x_d) \ee
where $B$ is the basis of mixed moments we are using for our modelling. It is required to contain $(0,\ldots,0)$ for normalization. Beside, there is a freedom of choosing this basis, what allows to hierarchically decompose statistical dependencies of multiple variables into mixed moments.

Figure \ref{intr} contains some first 5 functions of such product basis for $d=2$ variables: $f_{00}$ corresponds to normalization and requires $a_{00}=1$. Coefficients of $f_{10}$, $f_{20}$ describe expected value and variance of the first variable, $f_{01}$ and $f_{02}$ analogously of the second. Then we can start including moment dependencies, starting with $a_{11}$ which determines decrease/increase of expected value of one variable with growth of expected value of the second variables - analogously to correlation coefficient. We have also dependencies between higher moments, like asymmetric $a_{12}$ relating expected value of the first variable and variance of the second.

And analogously for more variables, e.g. $a_{010010}$ describes correlation between 2nd and 5th out of 6 variables. Finally we can hierarchical decompose statistical dependencies between multiple variables into their mixed moments. However, to fully describe general joint distribution, we would need $B=\mathbb{N}^d$ infinite number of mixed moments this way - for practical applications we need to choose some finite basis $B$ of moments to focus on.

\subsection{Estimation with least squares linear regression}
Having a data sample $\mathcal{X}$, we would like to estimate such mixed moments as coefficients for linear combination of some orthonomal basis of functions e.g. polynomials. Smoothing the sample with kernel density estimation, finding linear combination minimizing square distance to such smoothened sample, and performing limit to zero width of the used kernel, we get convenient and inexpensive MSE estimation~\cite{HCR}: independently for each coefficient $j$ as just average over dataset of value of the corresponding function:
\be a_j= \frac{1}{|\mathcal{X}|} \sum_{x\in \mathcal{X}} f_j(x) \label{est} \ee
We could use such obtained model for predicting conditional distribution: substitute the known variables to the modeled joint distribution, after normalization getting (conditional) density of the unknown variables.

However, for the bid-ask spread prediction problem, slightly better evaluation was obtained by generalizing alternative approach from \cite{cred}, which allows to additionally exploit subtle variable dependencies, hence we will focus on it.

\begin{figure}[t!]
    \centering
        \includegraphics{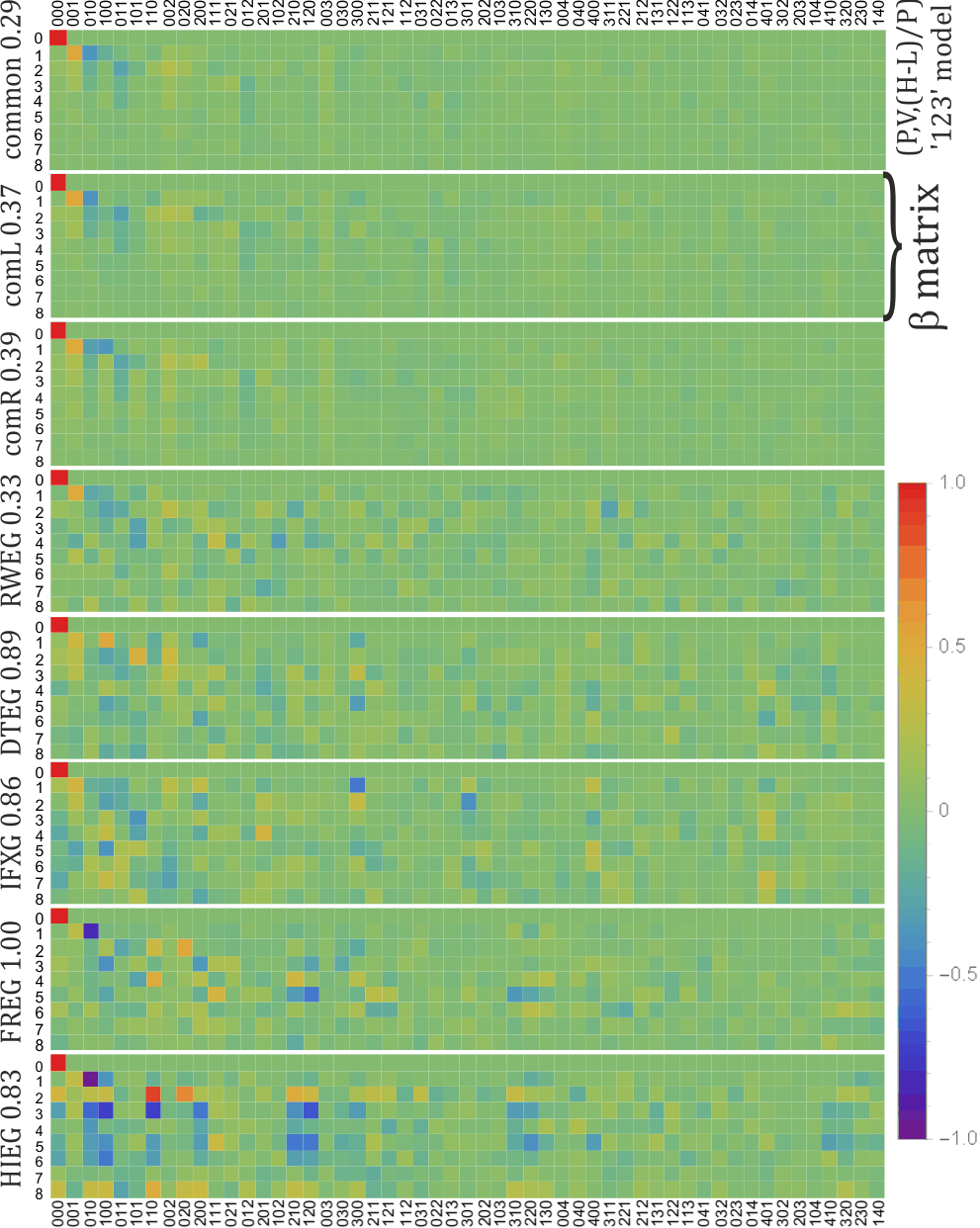}
        \caption{Visualized coefficients of '123' models ($9\times 53$ matrix $\beta$ for $\rho(y|x)= \sum_j f_j(y) \sum_k \beta_{jk} f_k(x)$), the numbers above names are log-likelihoods. The 'common' is the model built for combined all data - presents general trends. Trying to split all companies into subsets of similar behavior, as visualized in tree Fig. \ref{tree}, splitting into two subsets we get the presented comL and comR models correspondingly for left (DPWG, BEIG, HNKG, FME, SAPG, DB1G, RWEG, FREG, HEIG, DTEG, IFXG) and right (DAIG, SIEG, TKA, CONG, MRCG, LHAG, VOWG, MUVG, ALVG, BMW, DBKG) subtree of this tree. Then there presented individual models for 5 selected companies. Rows correspond to predicted moments of $Y$, as a linear combinations of mixed moments of $X$ corresponding to columns. The zeroth row has always only 000 nonzero coefficient equal 1 for normalization. The next row describes prediction of expected value, the next one of variance and so on. In the top model, common for all companies, we can e.g. see large positive $001\to 1$ coefficient: spread increases with growth of $H-L$, negative $010\to 1$: spread decreases with growth of $V$, and negative $011\to 2$: variance of spread decreases for correlated $V$ and $H-L$. Blue $100\to 3$ for FREG denotes reduction of skewness of spread with growth of price. Generally, we can see quite individual behavior for different companies, starting with $100\to 1$ analogous to price-spread correlation, which seems the main dividing factor between comL and comR companies. }
       \label{models}
\end{figure}

Specifically, to model $\rho(Y=y|X=x)$, let us use separate bases of (mixed) moments: $B_X$ for $X$, $B_Y$ for $Y$, and model relations between them. While there could be considered more sophisticated models for such relations including neural networks, for simplicity and interpretability we focus on linear models here, treating $f_j(x)$ as interpretable features:
\be \rho(y|x) = \sum_{j\in  B_Y} f_j(y) a_j(x) \quad\textrm{for}\quad a_j(x) = \sum_{k\in B_X} \beta_{jk} f_k(x) \label{mod1}\ee
hence the model is defined by the $\beta$ matrix, which examples are visualized in Fig. \ref{models} for $|B_X|=53$, $|B_Y|=1+8$. 

It allows for good interpretability: $\beta_{jk}$ coefficient is linear contribution of $k$-th mixed moment of $X$ to $j$-th (mixed) moment of $Y$. We focus on one-dimensional $Y$, but the formalism allows to analogously predict multidimensional.

To find the $\beta$ model we use least-squares optimization here - it is very inexpensive, can be made independently for each $j\in B_Y$ thanks to using orthonormal basis, and intuitively is a proper heuristic: least-squares optimization estimates the mean - exactly as we would like for coefficient estimation (\ref{est}). However, this is not necessarily the optimal choice - it might be worth to  explore also more sophisticated ways.

Such least-squares optimization has to be performed separately for each $j\in B_Y$. Denoting $\beta_{j\boldsymbol{\cdot}}=(\beta_{jk})_{k\in B_X}$ as coefficient vector for $j$-th moment and $\mathcal{Z}=\{(y^i,x^i)\}_{i=1..n}$ as (e.g. training) dataset of $(y,x)$ pairs:
$$\beta_{j\boldsymbol{\cdot}} = \textrm{argmin}_v \sum_{(y,x)\in \mathcal{Z}}  \left(\sum_{k\in B_X} f_k(x)v_k-f_j(y)\right)^2 =$$
$$=\textrm{argmin}_v \left\|Mv - b^j\right\|^2$$
$$\textrm{for } M = [f_k(x^i)]_{i=1..n,k\in B_X}\qquad b^j=(f_j(x^i))_{i=1..n}$$
matrix $M$ and vector $b^j$ for $j\in B_Y$. Such least-squares optimization has unique solution:
\be \beta_{j\boldsymbol{\cdot}} = (M^T M)^{-1} M^T b^j\ee
Separately calculated for each $j\in B_Y$, leading to the entire model as $\beta$ matrix with $\beta_{j\boldsymbol{\cdot}}$ rows, like in Fig. \ref{models}.
\subsection{Applying the model, enforcing nonnegativity}
Having such model $\beta$ we can apply it to (e.g. test) datapoints as in \ref{mod1}, getting predicted conditional density for $y$ on $[0,1]$ as a polynomial. However, sometimes it can get below 0, so let us refer to it as $\tilde{\rho}$ and then enforce nonnegativity required for densities:
\be \tilde{\rho}(y|x)= \sum_{j\in B_Y} f_j(y) \sum_{k\in B_X} \beta_{jk} f_k(x) \ee
Such obtained polynomial always integrates to 1. However, it occasionally can get below zero, what should be interpreted as corresponding to some low positive density. Such interpretation to nonnegative density $\rho$ is referred as calibration, and can be optimized based on dataset as discussed in \cite{hcr1}. For simplicity there was just used:
\be \rho(y|x)=\max\left(\tilde{\rho}(y|x),0.03\right)/N \ee
where $N$ normalization factor is chosen to integrate to 1: $N=\int_0^1 \max\left(\tilde{\rho}(y|x),0.03\right) dy$. The $0.03$ threshold was experimentally chosen as a compromise for the used dataset, its tuning can slightly improve evaluation.

Figure \ref{exdens} contains examples of such $\rho(y|x^i)$ predicted densities on the test set with $y^i$ actual values marked as vertical lines. Flat near zero regions come from $\max(\boldsymbol{\cdot},0.03)$ thresholding. While they are relatively frequent in such predicted densities, in plots of sorted $\{\rho(y^i|x^i)\}_i$ below we can see that these close to zero densities are very rare among the actual values: prediction properly excludes these $\tilde{\rho}<0.03$ regions as unlikely.

Integration is relatively costly to compute, especially in higher dimensions, hence for efficient calculation the predicted polynomial $\tilde{\rho}$ was discretized here into 100 values on $((i-0.5)/100)_{i=1..100}$ lattice, what corresponds to approximating density with piecewise constant function on length 1/100 subranges. Then $\max(\boldsymbol{\cdot},0.03)$ was applied, and division by sum for normalization. Finally density in discretized $\lceil 100y^i\rceil/100$ position was used as $\rho(y^i|x^i)$ in log-likelihood evaluation.
\subsection{Basic basis selection}
The optimal choice of basis is a difficult open question. As the basic choice there was used combinatorial family:
\be \mathcal{B}((m_1,\ldots,m_d),s,r):=\ee
$$=\left\{j\in \mathbb{N}^d: \forall_i j_i\leq m_i,\ \sum_{i=1}^d j_i \leq s,\ \sum_{i=1}^d \textrm{sgn}(j_i) \leq r\right\} \label{basis}$$
where $m_i$ chooses how many first moments to use for $i$-th variable, $s$ bounds the sum of used moments (and formally degree of corresponding polynomial), $r$ bounds the number of nonzero $j_i$: to include dependencies of up to $r$ variables.

For example the '123' model infers 8 moments $B_Y=\mathcal{B}((8),8,1)$ from 3 variables using a compromise: $B_X=\mathcal{B}((4,4,4),5,3)$ of size $|B_X|=53$ basis, directly written e.g. in Fig. \ref{models}. The '123+' model infers 6 moments $B_Y=\mathcal{B}((6),6,1)$ from 6 variables: $B_X=\mathcal{B}((4,4,3,1,2,1),5,3)$ of size $|B_X|=205$.
\section{Bid-ask spread modelling}
This section discusses application of the presented methodology to model conditional distribution of (relative quoted) bid-ask spread.
\begin{figure}[t!]
    \centering
        \includegraphics{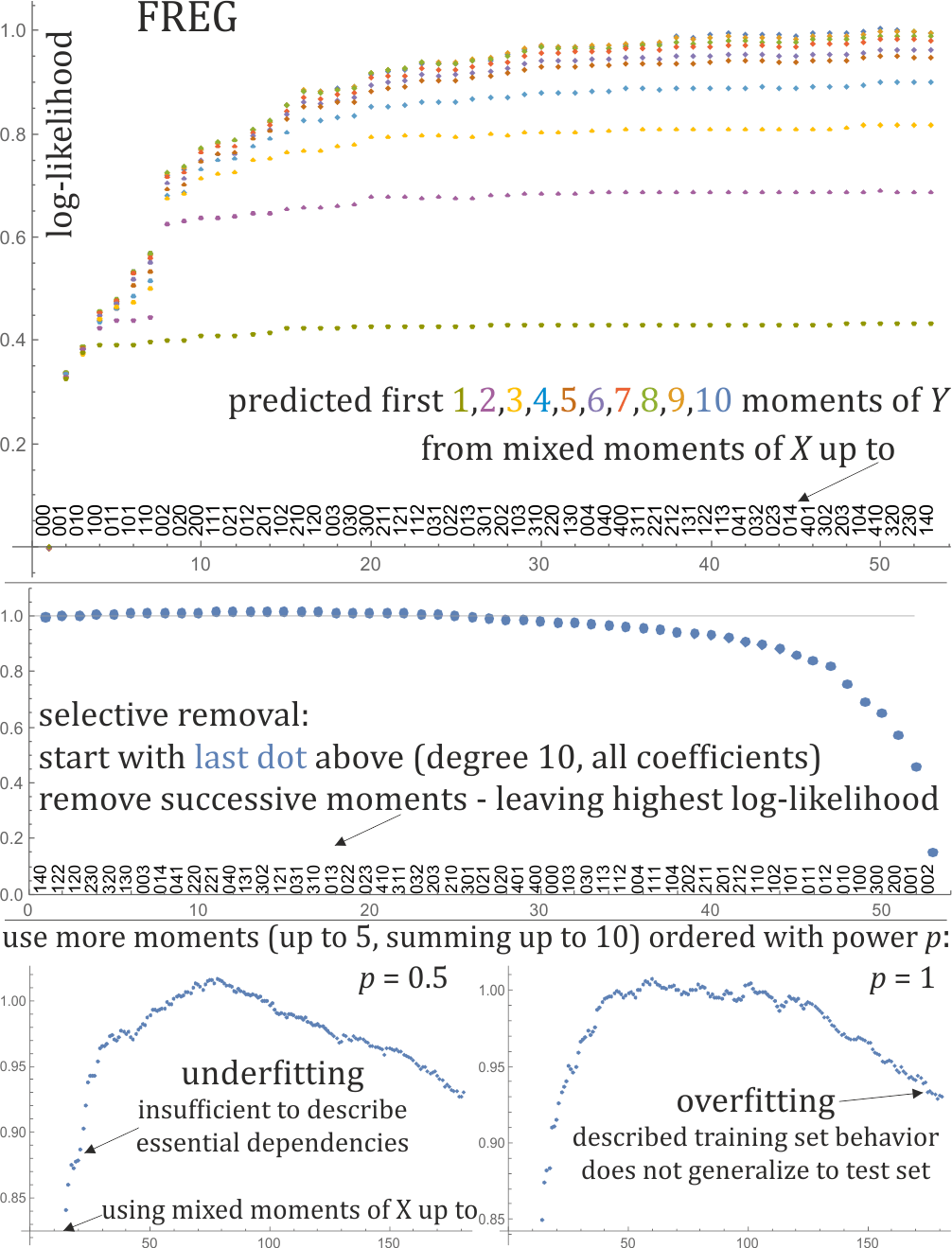}
        \caption{Top: Optimizing basis and model size on example of FREG company and $B_X=\mathcal{B}((4,4,4),5,3)$ size 53 basis of mixed moments from '123' model. Log-likelihoods for predicting first $1\ldots 10$ moments (denoted by colors) using some first of mixed moments (sorted lexicographically) of 3 $X$ variables: $P, V, (H-L)/P$. We can see that we should predict $\approx 8$ moments, higher moments are necessary to represent more complex distributions. Middle: selective removal of successive mixed moments to maximize log-likelihood - we can see that we can slightly improve evaluation this way, additionally reducing model size. However, it rather requires individual optimization for each company. Bottom: analogously as top, but using size 181 larger $B_X=\mathcal{B}((5,5,5),10,3)$, also trying different order of mixed moments: accordingly to $\sum_i (j_i)^p$. While using all such mixed moments clearly leads to overfitting, selectively using some first of them can lead to slightly improved evaluation.}
       \label{modelsize}
\end{figure}
\subsection{'123' model using basic variables}
The initial plan for this article was to improve prediction from standard models:  $AMI, HLR$, trying to predict conditional distribution of spread from their values using the discussed methodology. However, the results were disappointing, especially for $AMI$, as we can see in Fig. \ref{comp}.

Therefore, we have decided to use the original variables ($P,V,L,H,R$) instead, what has turned out to lead to essentially better predictions. There was manually performed search for  parameters using $\mathcal{B}$ basic basis selection (\ref{basis}) to maximize averaged log-likelihood in 10-fold cross validation. This search has finally lead to $B_X=\mathcal{B}((4,4,4),5,3)$ basis for only 3 variables: $P,V,(H-L)/P$ to predict up to 8-th moment of $Y$. Surprisingly, adding dependence on $R$ and $L$ alone was worsening evaluation - their dependence did not generalize from training to test sets.

While the optimal choice of basis seems a difficult open problem, exhaustive search over all subsets is rather impractically costly, Fig. \ref{modelsize} presents some heuristic approaches. The $\mathcal{B}$ family seems generally a good start, e.g. successively modifying some parameter by one as long as observing improvement. In this Figure we can see large improvement while rising the number of predicted moments up to $\approx 7$, what suggests that complexity of conditional distributions for the considered problem requires this degree of polynomial for proper description. This Figure also contains trials of using some first of such mixed moments accordingly to different orders. A heuristic optimization of a reasonable cost is the presented selective removal: for each mixed moment from $B_X$ calculate evaluation when it is removed, finally remove the one leading to the best evaluation, and so on as long as evaluation improves.

\subsection{Individual vs common models, universality}
A natural question is how helpful for prediction is a given variable - Figure \ref{lls} presents some answers by calculating log-likelihood also for models using only some of the variables. We can see different companies can have very different behavior here, e.g. for some $V$ is helpful, for some it is not, what we can also see in the presented points from dataset. Figure \ref{models} shows that they can even have opposite behavior: e.g. $100\to 1$ dependence on price.

This is a general lesson that while we would like predictors as nice simple formulas, the reality might be much more complicated - models found here are results of cultures of traders of stocks of individual companies, which can  essentially vary between companies.

Therefore, to get the most accurate predictions we should build individual models for each company. Even more, a specific behavior of a given company can additionally evolve in time - what could be exploited e.g. by building separate models for shorter time periods, or using adaptive least-squares linear regression \cite{adapt}, and is planned for future investigation.

\begin{figure}[t!]
    \centering
        \includegraphics{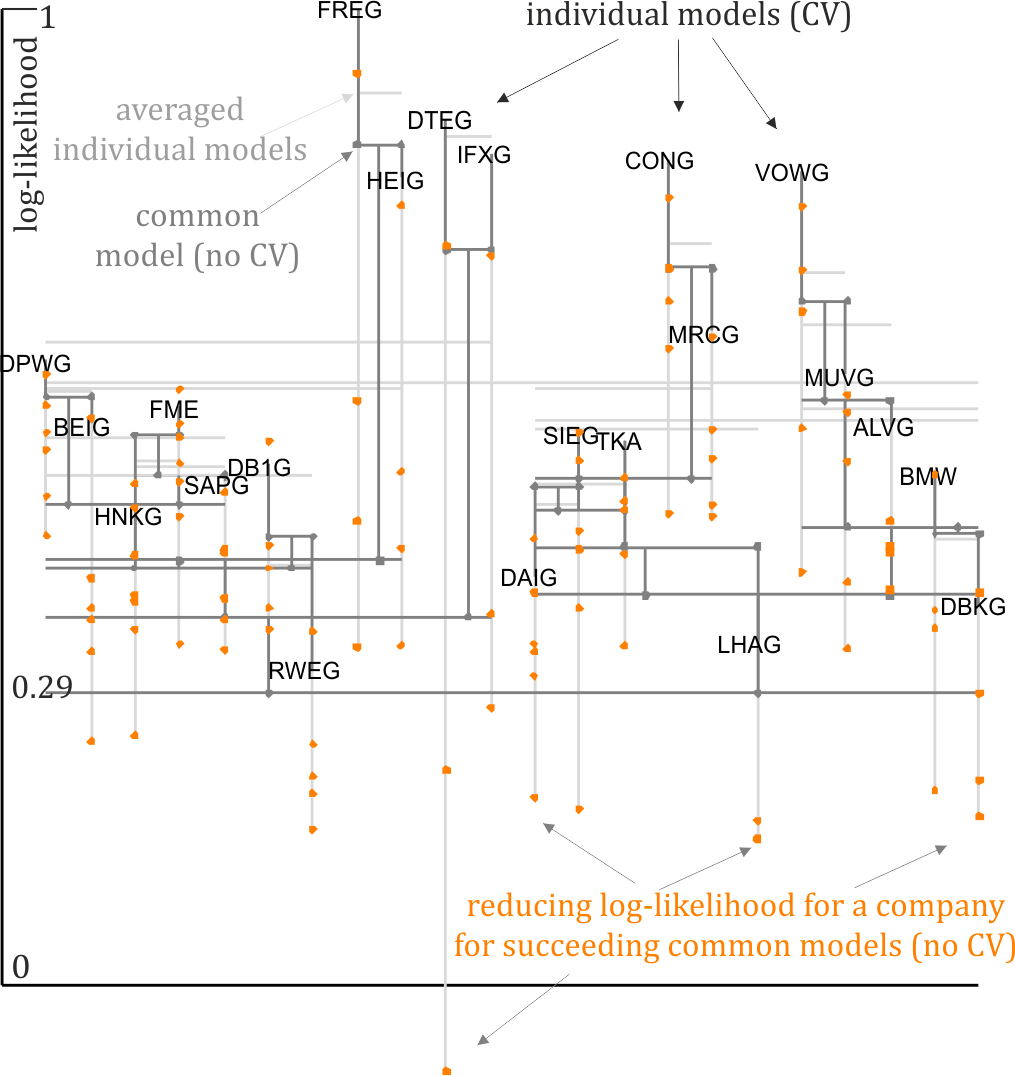}
        \caption{Visualization of optimized hierarchical grouping and loss of  using common models for multiple companies, height denotes log-likelihoods. Heights of names show evaluation of using individual model for a given company, orange dots show successive reduction of log-likelihood for a given company while using common models for subsets growing accordingly to the presented tree. The lowest dots correspond to using one common model for all (common in Fig. \ref{models})  - we can see that only for DTEG it is worse than zero (using no prediction at all). Splitting companies into left and right subtree and using separate two models for them (comL and comR in Fig. \ref{models}), we get essentially better prediction (one dot up). The tree structure was calculated by combining subsets to maximize (log-likelihood of common model / average log-likelihood of individual models) - grouping companies into pairs and then further, up to a single common model for all. Positions of lines represent such grouped companies: light-gray line their averaged log-likelihoods of individual models, dark-gray line their log-likelihood for a common model. The difference between these two lines represent loss of using common model. The common models are fixed hence there is no cross validation (CV), what artificially improves performance, for example for the first dot of FME corresponding to common model with HNKG - making it above CV individual model, generally suggesting large time inhomogeneities - to be included in future adaptive models.}
       \label{tree}
\end{figure}

However, building such models requires training data, which in case of variables like bid-ask spread might be difficult to access. Hence it is also important to search for universality - e.g. try to guess a model for a company for which we lack such data, based on data available for other companies. This generally seems a very difficult problem, Fig. \ref{tree} shows that even having all the data, using common model for multiple companies we should expect large evaluation drop. For example we can see that behavior of DTEG completely disagrees with common model for all.

As we can see in this tree Figure, the one common model situation improves if we can cluster companies into groups of similar behavior (orange dots) - there are also presented results for splitting companies into just two groups with separate models (comL, comR in Fig. \ref{models}), also visually leading to slightly better prediction as we can see comparing 3rd and 4th column in Fig. \ref{comp}.

\subsection{'123+' model with additional variables}
The information from $P, V, H, L, R$ basic variables can often be complemented with some additional - a natural approach is checking if we can improve log-likelihood in the discussed methodology if adding information from some new variables.

The size of basis can even grow exponentially with the number of variables here: there are $(m+1)^d$ mixed moments if using all up to $m$-th moment for all $d$ variables. The $r$ parameter: maximal number of interacting variables in $\sum_{i=1}^d \textrm{sgn}(j_i) \leq r$ allows to bound it by $O((dm)^r)$. The sum $s$ limitation in $\sum_{i=1}^d j_i \leq s$ seems also very useful, bounding degree of used polynomials.

Due to quickly growing basis size for increasing number of variables, we could easily exceed the size of dataset - experimentally seen as overfitting: decreasing performance. Manual search for using additional variables has started with $B_X=\mathcal{B}((4,4,4),5,3)$ basis for the standard variables, and carefully increasing $m$ in $B_X=\mathcal{B}((4,4,4,m),5,3)$ basis with separate single additional variable to consider. The most promising variables were later considered together, by modifying parameters by 1 as long as improvement was observed (of averaged log-likelihood over individual models for all companies).

It has finally lead to '123+' model: $B_Y=\mathcal{B}((6),6,1)$ and $|B_X|=205$ size $B_X=\mathcal{B}((4,4,3,1,2,1),5,3)$ using 3 additional variables: depth, midpoint changes intraday and midpoint volatility. Such model has $6\cdot 205 = 1230$ coefficients.

Due to rapid growth of the number of coefficients, for adding further variables it is worth to consider e.g. building some features from multiple variables to be directly used here, or use some alternative way for choosing basis for $X$, e.g. directly optimized on the dataset like PCA or other dimensionality reduction.

\section{Conclusions and further work}
There was presented a general methodology for extracting and exploiting complex statistical dependencies between multiple variables in inexpensive and interpretable way for predicting conditional probability distributions, on example of difficult problem of predicting bid-ask spread from more available information. It expands approach form \cite{cred} by inferring from mixed moments, and searching for the basis in large spaces of possibilities.

\begin{figure*}[b!]
    \centering
        \includegraphics{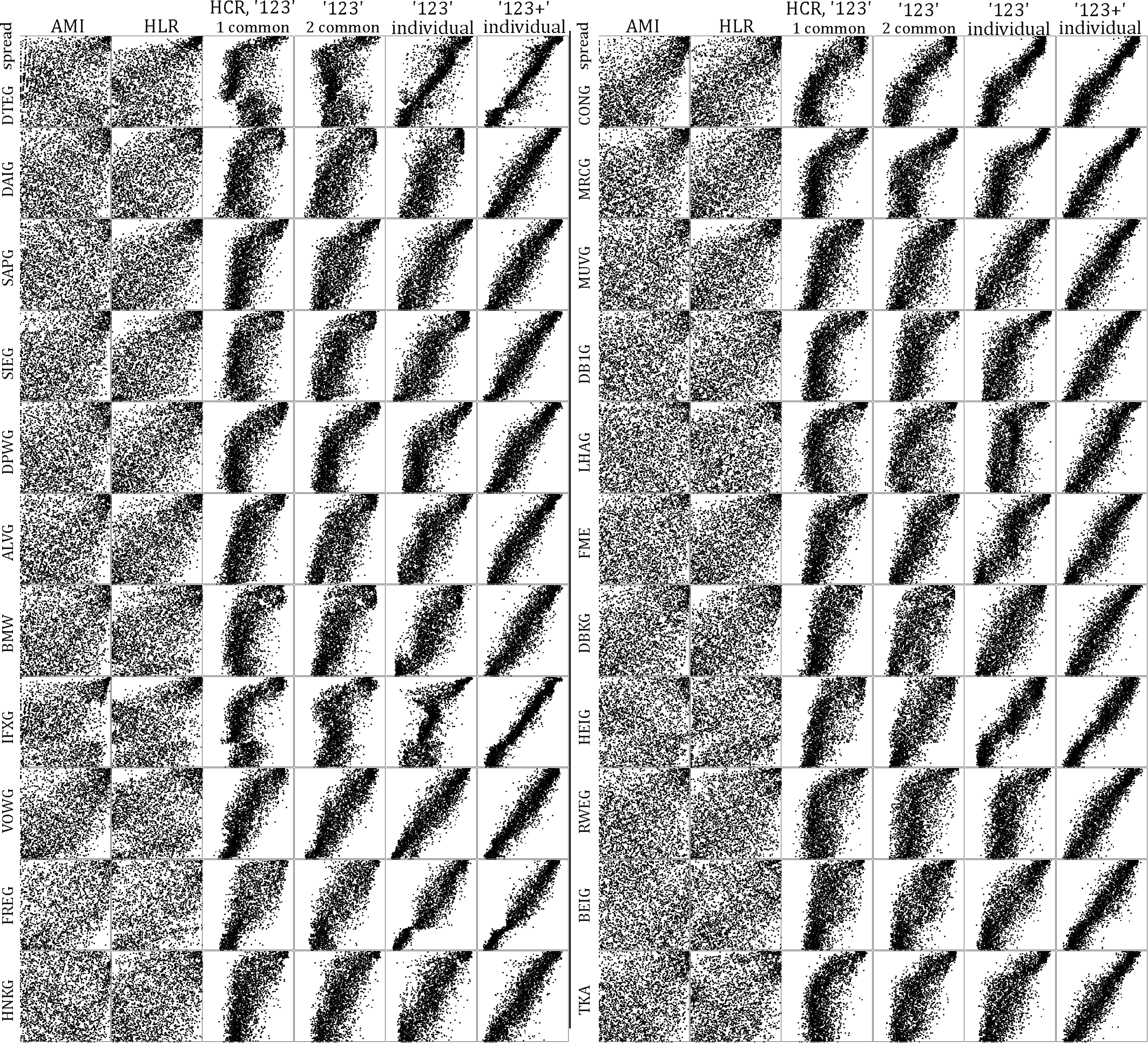}
        \caption{Comparison of spread predictors on dataset for visual evaluation: perfect predictor would give a diagonal, a completely useless one would give uniform distribution. All variables are normalized to nearly uniform marginal distributions, including outcomes of standard methods: $AMI$, $HLR$. The following 3 columns use expected values of predicted densities from discussed '123' model (using $P, V, (H-L)/P$ variables, $8\cdot 53=424$ coefficients), the last one is for '123+' model (using also depth, midpoint changes intraday and midpoint volatility,  $6\times 205=1230$ coefficients). The "1 common" column uses 1 model for all, "2 common" groups companies into two subsets and uses one of two models (as in Fig. \ref{tree}, using models comL, comR from Fig. \ref{models}). The last two columns use models individually optimized for each company.}
       \label{comp}
\end{figure*}

Figure \ref{comp} presents its comparison with standard methods when using only expected value from such predicted conditional density - perfect predictor would lead to diagonal, standard methods give rather a noise instead, while the predictions from the discussed approaches indeed often resemble diagonal, especially when using individual models. Predicted conditional probability density provides much more information: e.g. allows to additionally estimate uncertainty of such prediction, or provide or-or prediction for multimodal densities, or allows for generating its random values e.g. for Monte-Carlo simulations, or just provide the entire density for accurate considerations if transforming such random variables through some further functions.\\

There are many directions for further development of this relatively new general methodology, for example:
\begin{itemize}
  \item Optimal choice of basis is a difficult problem, necessary to be automatized especially for a larger number of variables - selecting from discussed basis of orthonormal polynomials, or maybe automatically optimizing a completely different basis based on dataset.
  \item There are observed large differences between behaviors of individual companies - bringing difficult questions of trying to optimize for common behavior, optimize models based on incomplete information, etc. Additionally, such behavior has probably also time inhomogeneity - the models should evolve in time, requiring adaptive models to improve performance, where the problem of data availability becomes even more crucial.
  \item The discussed models rapidly grow with the number of variables, what requires some modifications for exploiting high dimensional information - like extracting features from these variables, dimensionality reduction like PCA, etc.
  \item We have predicted conditional distribution for one-dimensional variable, but the methodology was introduced to be more general: predicting for multidimensional $Y$ should be just a matter of using proper $B_Y$, what is planned to be tested in future.
  \item The predicted densities as polynomials have often rapid growths at the ends of $[0,1]$ - their removal might improve performance.
  \item There was assumed linear relation between moments with least-squares optimization, what is inexpensive and has good interpretability, but is not necessarily optimal - there could be considered e.g. using neural networks instead, and optimizing criteria closer to log-likelihood of final predictions.
\end{itemize}
\section*{Acknowledgement}
Henryk Gurgul  thanks Professor Roland Mestel for providing the bid-ask data from data bank "Finance Research Graz Data Services" and Professor Erik Theissen and  Stefan Scharnowski from Mannheim for providing data from  "Market Microstructure Database".

\bibliographystyle{IEEEtran}
\bibliography{cites}

\begin{thebibliography}{10}
\providecommand{\url}[1]{#1}
\csname url@samestyle\endcsname
\providecommand{\newblock}{\relax}
\providecommand{\bibinfo}[2]{#2}
\providecommand{\BIBentrySTDinterwordspacing}{\spaceskip=0pt\relax}
\providecommand{\BIBentryALTinterwordstretchfactor}{4}
\providecommand{\BIBentryALTinterwordspacing}{\spaceskip=\fontdimen2\font plus
\BIBentryALTinterwordstretchfactor\fontdimen3\font minus
  \fontdimen4\font\relax}
\providecommand{\BIBforeignlanguage}[2]{{%
\expandafter\ifx\csname l@#1\endcsname\relax
\typeout{** WARNING: IEEEtran.bst: No hyphenation pattern has been}%
\typeout{** loaded for the language `#1'. Using the pattern for}%
\typeout{** the default language instead.}%
\else
\language=\csname l@#1\endcsname
\fi
#2}}
\providecommand{\BIBdecl}{\relax}
\BIBdecl

\bibitem{HCR}
J.~Duda, ``Hierarchical correlation reconstruction with missing data, for
  example for biology-inspired neuron,'' \emph{arXiv preprint
  arXiv:1804.06218}, 2018.

\bibitem{copula}
F.~Durante and C.~Sempi, \emph{Principles of copula theory}.\hskip 1em plus
  0.5em minus 0.4em\relax Chapman and Hall/CRC, 2015.

\bibitem{cred}
J.~Duda and A.~Szulc, ``Credibility evaluation of income data with hierarchical
  correlation reconstruction,'' \emph{arXiv preprint arXiv:1812.08040}, 2018.

\bibitem{liq1}
R.~Mestel, M.~Murg, and E.~Theissen, ``Algorithmic trading and liquidity: Long
  term evidence from austria,'' \emph{Finance Research Letters}, vol.~26, pp.
  198--203, 2018.

\bibitem{liq2}
H.~Gurgul and A.~Machno, ``The impact of asynchronous trading on epps effect on
  warsaw stock exchange,'' \emph{Central European Journal of Operations
  Research}, vol.~25, no.~2, pp. 287--301, 2017.

\bibitem{AMI}
Y.~Amihud, ``Illiquidity and stock returns: cross-section and time-series
  effects,'' \emph{Journal of financial markets}, vol.~5, no.~1, pp. 31--56,
  2002.

\bibitem{AMI1}
K.~Y. Fong, C.~W. Holden, and C.~A. Trzcinka, ``What are the best liquidity
  proxies for global research?'' \emph{Review of Finance}, vol.~21, no.~4, pp.
  1355--1401, 2017.

\bibitem{HLR}
B.~Bedowska-S{\'o}jka and K.~Echaust, ``Commonality in liquidity indices: The
  emerging european stock markets,'' \emph{Systems}, vol.~7, no.~2, p.~24,
  2019.

\bibitem{hcr1}
J.~Duda, ``Exploiting statistical dependencies of time series with hierarchical
  correlation reconstruction,'' \emph{arXiv preprint arXiv:1807.04119}, 2018.

\bibitem{adapt}
------, ``Parametric context adaptive laplace distribution for multimedia
  compression,'' \emph{arXiv preprint arXiv:1906.03238}, 2019.

\end{thebibliography}
\end{document}